# Overcoming losses in superlenses with synthetic waves of complex frequency

Fuxin Guan[1], Kebo Zeng[1], Zhaoyu Nie[2], Xiangdong Guo[3], Shaojie Ma[1], Qing Dai[3], John Pendry[4,*], Xiang Zhang[1,5,*] and Shuang Zhang[1,6,*]

[1]Department of Physics, University of Hong Kong, Hong Kong, China

[2]Department of Mechanical Engineering, University of California, Berkeley, California 94720, United States

[3]Center of Materials Science and Optoelectronics Engineering University of Chinese Academy of Sciences Beijing 100049, China

[4]The Blackett Laboratory, Department of Physics, Imperial College London, London SW7 2AZ, UK

[5]Faculty of Science and Faculty of Engineering, University of Hong Kong, Hong Kong, China

[6]Department of Electrical & Electronic Engineering, University of Hong Kong, Hong Kong, China

[*]Corresponding author: j.pendry@imperial.ac.uk, president@hku.edu, shuzhang@hku.hk

Superlenses made of plasmonic materials and metamaterials have been exploited to image features of sub-diffractional scale. However, their intrinsic losses impose a serious restriction on the imaging resolution, which is a long-standing problem that has hindered wide-spread applications of superlenses. Optical waves of complex frequency exhibiting a temporally attenuating behavior have been proposed to offset the intrinsic losses in superlenses via virtual gain, but the experimental realization has been missing due to the challenge involved in preparing the illumination with temporal decay. Here, by employing multi-frequency measurement, we successfully implement a synthetic optical wave of complex frequency to experimentally observe deep-subwavelength superimaging patterns enabled by the virtual gain. Our work represents a practical approach to overcoming the intrinsic losses of plasmonic systems for imaging and sensing applications.

Abbe diffraction limits the conventional optical imaging resolution to larger than half the wavelength due to the missing of the subwavelength information carried by evanescent waves. To overcome this limitation, negative refractive index lens has been proposed to significantly enhance the evanescent waves to recover the deep-subwavelength resolution of imaging [1,2]. Subsequently, superlenses, made of either natural materials with negative permittivities [3–6], or hyperbolic materials with mixed signs of dielectric constants along different directions [7,8], have been proposed to attain sub-diffractional limited imaging. Nevertheless, losses are non-negligible in materials with negative parameters [9–11], which significantly reduces the deep-subwavelength information of the superlenses and seriously affects the resolution of imaging [12–14]. To compensate the losses, it has been proposed that gain materials could be incorporated into metamaterial designs or plasmonics [15–22], but the setup is extremely complicated and the gain will inevitably introduce instability and noise into the system [23–25]. Recently, complex-frequency waves with temporally growing or attenuating behaviors have been treated as virtual absorption or gain waves [26]. Some theoretical proposals have been put forward to recover the deep-subwavelength information carried by surface plasmons via excitation of complex-frequency waves (CFW) with temporal attenuation [27,28]. However, synthesizing CFW is a significant challenge in optical systems from a practical perspective, and superimaging with CFW has not been experimentally achieved.

Here, to address this challenge, we synthesize CFW signals using a multi-frequency approach. We exploit the fact that a truncated CFW can be expressed as combination of multiple frequency components following a Lorentzian lineshape in the frequency domain through the Fourier transformation. As a proof of the concept, a bulk hyperbolic metamaterial is employed as an imaging lens operating at microwave frequencies. We show that, while the spatial resolution of imaging at real frequencies is poor, caused by the inevitable material loss in the system, an ultra-high resolution imaging can be obtained via the synthesized complex frequency by using multiple frequency components.

We start with an example of loss compensation for a metallic material described by the Drude model, $\varepsilon(\omega) = 1 - \omega_p^2/(\omega^2 + i\omega\gamma)$, where $\gamma$ is the nonzero ohmic loss term. Below the plasma frequency $\omega_p$, the permittivity becomes negative, making it suitable as a plasmonic material, or for constructing hyperbolic media, to support surface or bulk waves with very large wavevector for sub-diffractional imaging. Due to the existence of loss term, the negative permittivity is typically accompanied by a significant imaginary part (left panel of Fig. 1(a)), which seriously limits the imaging performance. Interestingly, from a mathematical perspective, by transforming the frequency into the suitable complex value $\omega \to \omega - i\gamma/2$, the permittivity is turned into a purely real value $\varepsilon(\omega - i\gamma/2) = 1 - \omega_p^2/(\omega^2 + \gamma^2/4)$.

A CFW with negative imaginary part corresponds to a wave with temporal attenuation. The mathematical expression of a CFW is expressed as, $E(t) \sim e^{-i\tilde{\omega}t}$, where $t$ denotes time, $\tilde{\omega} = \omega - i\tau/2$, and $\tau > 0$ is the temporal attenuation factor. Although an ideal CFW exists mathematically, it is unphysical as it implies the energy would diverge when $t$ approaches negative infinity. Hence a truncation at the start of time needs to be implemented to rationalize the CFW, i.e. $E_T(t) \sim E_0 e^{-i\tilde{\omega}t}\theta(t)$, where $\theta(t) = 0$ for $t < 0$, and $\theta(t) = 1$ for $t \geq 0$. The constitutive equation of a dispersive medium under the truncated CFW, i.e. $D(t) = \int \varepsilon(t-t')E_T(t')dt'$, can be expressed as,

$$D(t) = \int \varepsilon(\omega')f(\tilde{\omega},\omega')e^{-i\omega't}d\omega'/2\pi \tag{1}$$

where $\omega'$ denotes the real frequency and $f(\tilde{\omega},\omega') = -i/(\tilde{\omega} - \omega')$ has the Lorentzian lineshape. Suppose the temporal duration is long enough, Eq. (1) can be further simplified according to residue theorem as, $D(t) = \varepsilon(\tilde{\omega})E_0 e^{-i\tilde{\omega}t}$. A rigorous proof of this formula is presented in the supplementary material. According to Eq. (1), any response $F(\tilde{\omega})$ of the system with complex frequency can be obtained via the integral of the real frequency responses $F(\omega')$, $F(\tilde{\omega}) \approx \int F(\omega')f(\tilde{\omega},\omega')e^{-i\omega't+i\tilde{\omega}t}d\omega'/2\pi$. In practice, it is sufficient to choose a number of discrete frequency points at a certain frequency interval $\Delta\omega$ to synthesize the signal according to Nyquist-Shannon sampling theorem. Further, based on the compressed sensing theory [29–31], the CFW can be synthesized based on the information taken from a finite spectrum range. If the spectrum range is broad enough, the noise of the interference between different harmonics is suppressed. The response under synthesized CFW can be discretized as,

$$F(\widetilde{\omega}) \approx \sum_i F(\omega_i')f(\widetilde{\omega},\omega_i')e^{-i\omega_i't+i\widetilde{\omega}t}\Delta\omega'/2\pi \tag{2}$$

Since the frequency is discretized, the signal has an overall $2\pi/\Delta\omega'$ temporal periodicity. Feeding the permittivity at a number of frequencies (the black and grey circles in the left panel of Fig. 1(a)) into Eq. (2), we can obtain the synthesized permittivity of complex frequency with $\tau = \gamma$ as shown in the right panel of Fig. 1(a), which clearly shows that the loss of the Drude model can be largely compensated due to virtual gain. It should be noted that if the virtual gain of CFW is greater than the actual material loss $\tau > \gamma$, oscillatory behavior will arise, leading to an unstable state for the whole system.

We use the synthetic CFW to study the imaging performance in a plasmon metal/dielectric multilayer lens with a total of 15 layers as shown in Fig. 1(b), which can functions as a type II hyperbolic media with mixed signs of permittivity tensor elements along different directions [32]. The left panel of Fig. 1(b) schematically shows the field emitted from two closely spaced slits from the bottom of a flat hyperbolic material. The waves transmitted to the upper interface experience total internal reflection due to the large in-plane wavevectors, and form an electric field pattern with evanescent wave $E_{obj}(\omega,r)$ on the air side. The corresponding distribution in the momentum space $E_{obj}(\omega,k)$ can be derived via Fourier transformation, here *k* is the in-plane wavevector. Inspired by the time-reversal imaging technique that has been demonstrated in acoustics and other wave systems [33–39], we use a postprocessing to mimick a phase conjugation action to restore the image of the object. We first perform the complex conjugate operation of the momentum space distribution $E_{obj}(\omega,k) \to E_{obj}^*(\omega,k)$, then multiply it by the transfer function to obtain the image in the Fourier space, i.e.,

$$E_{imag}(\omega,k) = E_{obj}^*(\omega,k)t(\omega,k) \tag{3}$$

In the above equation, the transfer function is obtained via Fourier transforming the point spread function $t(\omega, k) = \mathcal{F}(E_{PSF}(\omega, r))$, where $E_{PSF}(\omega, r)$ is the field emitted from a point source located on one side of the hyperbolic slab to the opposite surface, as shown in right panel of Fig. 1(b). Finally, the real space image pattern $E_{imag}(\omega, r)$ can be obtained via the inverse Fourier transformation of Eq. (3).

We use FEM simulation to numerically calculate $E_{obj}^*(\omega, k)$ and $t(\omega, k)$, which are displayed in the left and right panels of Fig. 1(c), respectively. The calculated image pattern based on Eq. (3) at $f = 6.68\ GHz$ is shown by the red line in Fig. 1(e), which deviates significantly from the object (shown as the dashed line). In contrast, by performing phase conjugation in the complex frequency domain [40–42], i.e. by constructing the $E_{obj}^*(\widetilde{\omega}, k)$ and $t(\widetilde{\omega}, k)$ respectively using Eq. (2) with $\tau = \gamma$, the large wavevector components are recovered as shown in Fig. 1(d). The complex-frequency imaging pattern faithfully follows the original pattern as shown by the blue line in Fig. 1(e), verifying the capability of the multi-frequency approach to synthesize the CFW for enhancing the resolution.

Eqs. (2) and (3) require the distribution of both amplitude and phase of field, which can be readily obtained using a microwave characterization setup for a flat hyperbolic metamaterial lens designed to operate. The unit cell of the metamaterial consists of a spiral metallic wire/dielectric layer with $4mm \times 4mm \times 1.5mm$ to form a type-II hyperbolic metamaterial with two identical in-plane negative permittivities and one out-plane positive permittivity as shown in Fig. 2(a). The spiral structure can significantly reduce the plasma frequency such that the accessible wavevectors in the Brillouin zone can be much larger than that in the free space. The corresponding equi-frequency contours (EFCs) at different frequencies are retrieved using full-wave simulations as shown in Fig. 2(b), where the frequencies of the EFCs increase along the direction of the arrow. At higher frequencies, by neglecting the effect of loss, the EFC can reach the horizontal Brillouin edge, providing large in-plane wavevectors which are essential for achieving sub-diffractional imaging resolution. Fig. 2(c) displays the experimental setup, wherein the bulk metamaterial lens is composed of $80 \times 80$ in-plane unit cells and 25 layers vertically. A dipole source is placed at the bottom of the sample and a probe antenna is raster scanned at the top surface to measure the near field distribution.

To begin, we scan a 1D line above the sample to measure the field distribution, as depicted in Fig. 2(d), emitted from a single dipole source across 251 discrete frequency points within the range of 5GHz to 7.5GHz. The dispersion is subsequently obtained via Fourier transformation, as shown in Fig. 2(e). The dispersion plot exhibits two bright lines in the middle, which represent the light cone, while the other bright lines correspond to the hyperbolic modes. Limited by the damping of the system, the measured Fourier components in the momentum space are far from reaching the Brillouin zone edge. The corresponding effective electromagnetic parameters including the damping term can be obtained through the retrieval method based on the experimental field distribution [43], with the details shown in the supplementary material. As the frequency marked by the white dashed line (*f=6.5GHz*) exhibits relatively large wavevectors, it is selected as the central frequency for the signal synthesis. The 2D field distribution in the momentum space at the central frequency is shown in Fig. 2(f), and the distributions at other frequencies are provided in the supplementary material. Using Eq. (2), we synthesize the dispersion of the complex frequency using 251 frequency points, with the temporal snapshot captured at the end of the temporal periodicity, as shown in Fig. 2(g). Remarkably, the synthesized results with complex-frequency excitation recover the field components across a large portion of the Brillouin zone, resulting in a much better spatial resolution than that of the real frequency shown in Fig. 2(f). Note that the Fourier space distributions shown in Figs.2(f) and 2(g) correspond to the $t(\omega, k)$ and $t(\widetilde{\omega}, k)$ of real frequency and complex-frequency cases in Eq. (3), respectively.

We next showcase super-resolution imaging of subwavelength patterns with the synthesized complex frequency. The desired pattern is formed by sequentially placing a single emitting dipole antenna at different locations, and the field distribution at the top surface is measured at each location of the dipole source. The measured field distributions are then linearly superimposed, and a Fourier transformation is performed to obtain the Fourier patterns $E^*_{obj}(\omega, k)$ and $E^*_{obj}(\widetilde{\omega}, k)$. The ground truth of three different patterns, letter 'H', 'K' and 'U' are shown in Fig. 2(h), the corresponding imaging results obtained by using Eq. (3) are displayed in Figs. 2(i), 2(j), and 2(k), respectively. The top left of each subpanel shows the Fourier pattern with synthesized complex-frequency excitation, which occupies a significant portion of the Brillouin zone, while the corresponding left bottom of each subpanel displays the image of the subwavelength letters. In contrast, the imaging results of the central frequency are shown in the right subpanels for all three cases, and in each case, the original pattern cannot be identified, thereby confirming a substantial improvement in super-imaging with the multi-frequency approach.

We finally investigate the effect of the number of frequency points used to synthesize the complex-frequency response on the superimaging performance, with the results shown in Fig. 3. The upper-left subpanel of the figure displays the object, which is comprised of a 5×5 array of dipole antennas. The horizontal and vertical lattice constants are 1/4 and 1/6 of the central wavelength, respectively. We gradually reduce the number of frequency points from 251 to 1 to construct the synthetic imaging patterns, with a fixed frequency step of 0.01GHz, and the corresponding synthesized temporal signals are depicted in the insets. Our study shows that reducing the number of frequency points to 51 has a negligible impact on image quality. However, as we continue to reduce the number of frequency points, there is a significant decrease in imaging resolution, causing the images of the dipoles to merge into vertical lines. When the number of frequency points drops below 17, these lines become wider in the horizontal direction. This highlights the importance of having enough frequency points to maintain good spatial resolution and avoid image degradation.

We have presented a novel approach to compensate the intrinsic loss of a hyperbolic metamaterial superimaging system by synthesizing a complex-frequency excitation through a multi-frequency approach, which significantly improves imaging resolution beyond the limit imposed by the damping of the system. Our approach holds great potential for high-resolution microscopy. Furthermore, the synthesized complex-frequency approach can be extended to other areas of optics, such as plasmonic sensing applications. By leveraging the enhanced quality factor of plasmonic structures, our approach has the potential to significantly improve sensitivity in sensing applications. In addition, the approach can be tailored to different systems and geometries, providing a flexible and versatile tool for improving optical performance.

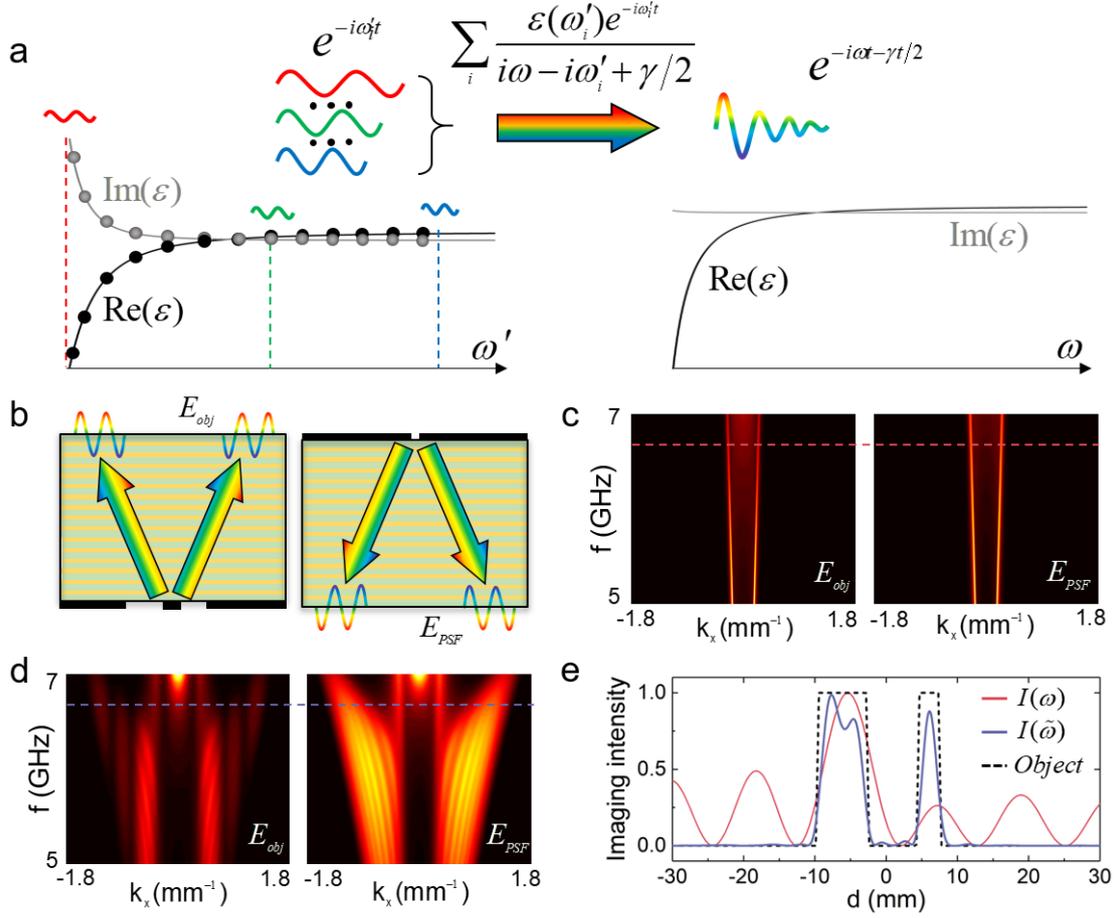

**Fig. 1. Illustration of loss compensation of superimaging lens using synthetic complex-frequency waves.** (a) The permittivities of the Drude model with a finite damping term before (left panel) and after (right panel) loss compensation using the complex-frequency waves synthesized with permittivities at multiple frequencies. The solid circles indicate the frequencies chosen to synthesize the complex-frequency permittivity. (b) Left panel shows a schematic of light passing through a plasmonic metal/dielectric multilayer lens from two closely placed slits at the bottom, where the thickness $d_m = 0.7mm$ for the Drude metal and $d_d = 1.76mm$ for the dielectric layer, respectively. The permittivity of metal is described by the Drude model with $\omega_p = 34\pi GHz, \gamma = 6.8\pi GHz$, and the dielectric one is $\varepsilon_d = 2.2$. $E_{obj}$ represents the field distribution at the top surface. The right panel shows the field pattern $E_{PSF}$ from a point source. (c) The dispersion plots formed by Fourier transformation of $E_{obj}$ and $E_{PSF}$ at different frequencies. (d) The complex-frequency dispersion plots derived by synthesizing the dispersions in (c) following Eq. (2). (e) The corresponding imaging

intensities at the positions of dashed lines in (c) and (d).

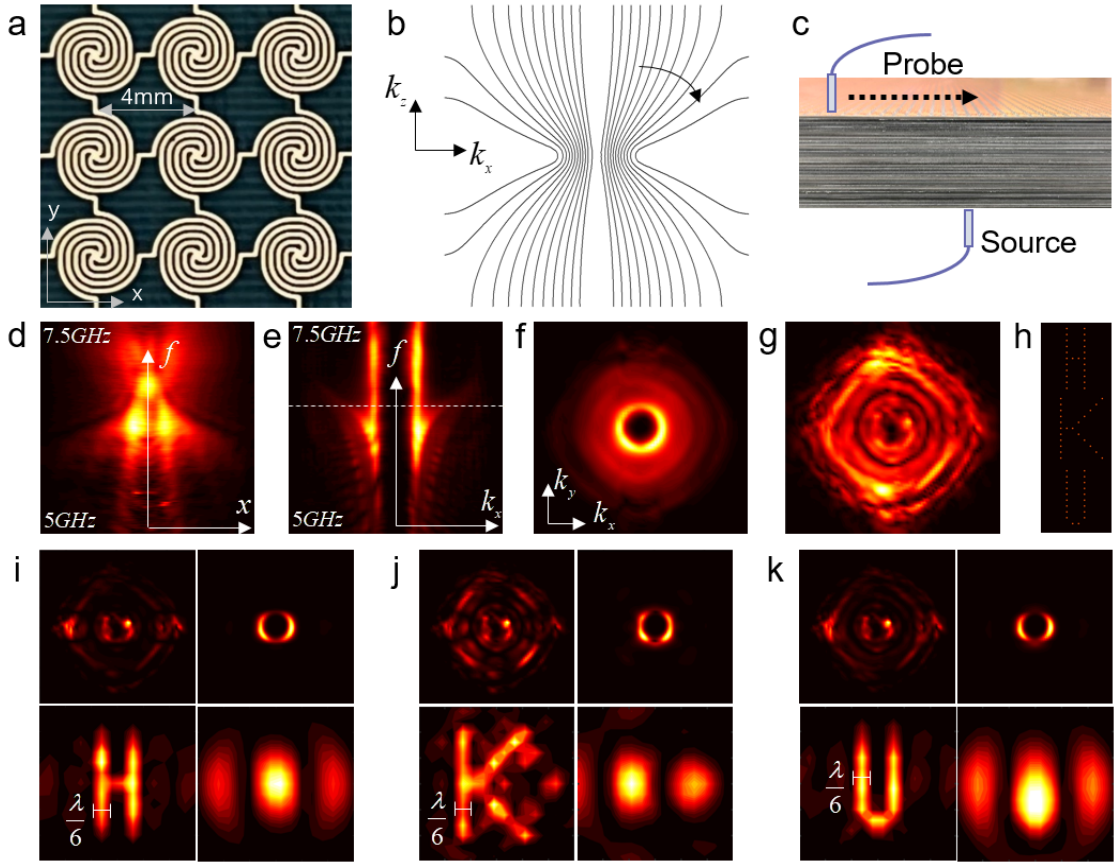

Fig. 2. **Experimental demonstration of loss compensation in superimaging using a hyperbolic lens at microwave frequency.** (a) Photograph of hyperbolic metamaterial with metallic spiral wires of 0.2mm width printed on a 1.5mm thick Teflon slab. The EFCs in the Brillouin zone at different frequencies are depicted in (b). The arrow implies the direction of increased frequency of the EFCs. (c) Schematic of the experimental setup with the two dipole antennas placed near the top and bottom interfaces of the sample, with the bottom and top one serving as source and probe, respectively. (d) The measured electric field distribution along a line at the top interface with a frequency step of 0.01GHz, which locates in the same y-z plane as the source antenna. (e) The dispersion plot obtained via spatial Fourier transformation of the real space field distribution shown in (d). (f) The 2D electric field distribution in the momentum space at a frequency of 6.5GHz, which serve as the transfer function. (g) The 2D electric field distribution in the momentum space for CFW synthesized by the measurements at multiple frequencies. (h) The ground truth of letters "H", "K", and "U". (i), (j), and (k) The imaging results with the dipole sources arranged in the shape

of the three letters. The upper and lower subpanels correspond to the Fourier space and real space results, and the left and right ones show the complex-frequency and real-frequency results, respectively.

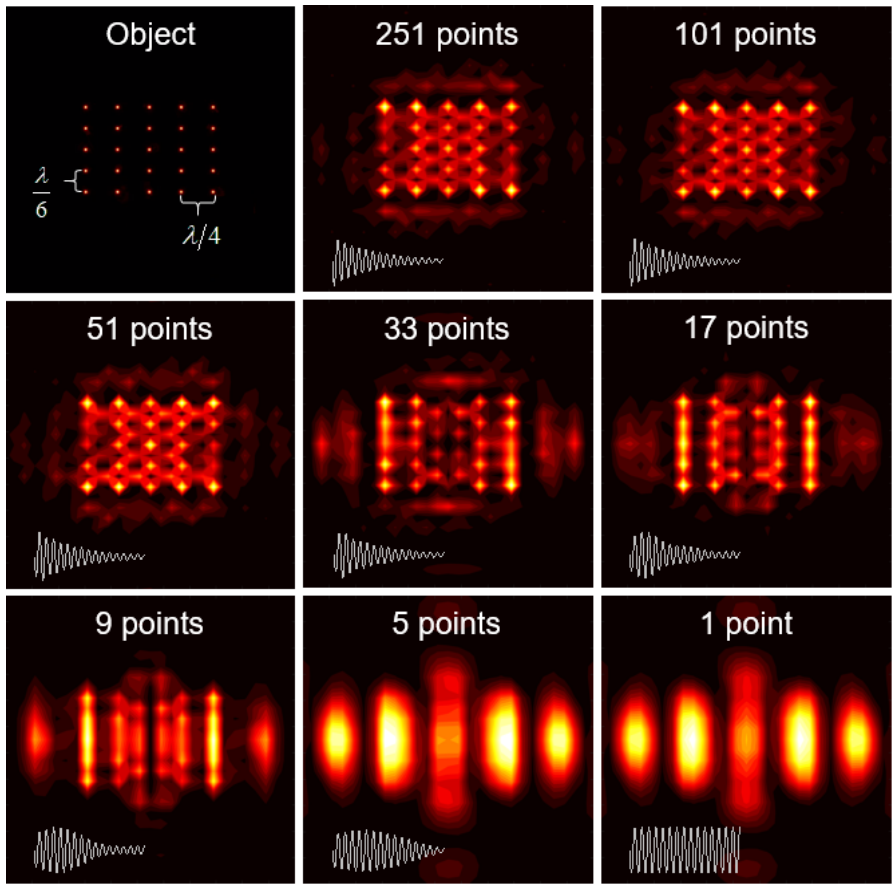

**Fig. 3. Investigation of the dependence of the imaging quality over the number of frequency points.** The ground truth is shown in the upper left, which consists of 5×5 antenna array, where the lattice constants along *x* direction and *y* direction are $\lambda/4$ and $\lambda/6$, respectively. The other subgraphs show the images obtained using different numbers of frequency points with the central frequency fixed at 6.5GHz, and a fixed frequency step of 0.01GHz. The corresponding synthetic temporal signal is depicted in the inset of each subpanel.